\documentclass[english]{article}
\usepackage[latin9]{inputenc}
\usepackage{geometry}
\usepackage{amsmath}
\usepackage{amssymb}
\usepackage{graphicx}

\makeatletter
\usepackage{babel}
\usepackage{placeins}
\usepackage{caption}

\makeatother

\usepackage{babel}
\begin{document}

\title{Matrix Models of Strength Distributions}

\author{Larry Zamick and Arun Kingan \\
 Department of Physics and Astronomy\\
 Rutgers University, Piscataway, New Jersey 08854 }
\maketitle
\begin{abstract}
In this work we use matrix models to study the problem of strength
distributions. This is motivated by noticing near exponential fall
offs of strengths in calculated magnetic dipole excitations. We emphasize
that the quality of the exponential fall offs depend on the parameters
in our matrices, especially the relative size of the couplings to
the unperturbed level separations. We also find a matrix for which
all transitions vanish.
 $ $\\
 PACs Number: 21.60.Cs 
\end{abstract}

\section{Introduction}

In this work we consider a matrix model of strength distributions.
Although the approach may seem somewhat abstract we should note that
the ideas came from very concrete calculations, especially calculations
of magnetic dipole transitions. An early experiment by Bohle et al.
{[}1{]} which found these low lying states in $^{196}$Gd lead to
a flood of papers (including some by one of us) but for brevity we
cite only the review article by Heyde et al. {[}2{]}. A more complete
list is given in the work of Harper and Zamick {[}3{]}. In references
{[}3,4 and 5{]} our group also focused, for the most part, on these
low lying excitations, but we also calculated the strength distributions
at higher energies. The results were rather messy but a crude examination
seemed to indicate an exponential decrease of the strength with excitation
energy. The ``mess'' was considerably reduced by a process called
binning {[}5{]}. Since we showed log(B(M1)) vs. excitation energy
we saw a linear behavior with a negative slope.

We here wish to address the same problem but using matrix models.
As is well known from the works of Heisenberg {[}6{]} and Born and
Jordan {[}7{]} the Hamiltonians can be represented by matrices. In
this work we will choose matrices with a somewhat simple structure
but ones that can still display complex behavior. By dealing with
matrices we can have better control and see how the distributions are
affected by the variations of the parameters.

\section{The model}

We represent a Hamiltonian by a matrix. On the diagonal we have what
can be considered unperturbed energy levels. We take them to be equally
spaced, E$_{n}$= n*E.

We introduce a constant coupling v which for a 
level E$_{n}$ occurs only with the nearest neighbors E$_{(n-1)}$ and
E$_{(n+1)}$. 

We consider a transition operator T such that the matrix 
element is non-zero only if n and n' differ by one and we take the 
non-vanishing matrix element to be a constant:

\bigskip

\textless{}\textless{}n T (n+1)\textgreater{}\textgreater{} =
\textless{}\textless{}(n+1) T n\textgreater{}\textgreater{} =
1 for all n. All other values are taken to be zero.

\section{The matrix}

The matrix that we will diagonalize is shown below.

\setcounter{MaxMatrixCols}{15}

\begin{gather*}
H=\begin{bmatrix}0 & v & 0 & 0 & 0 & 0 & 0 & 0 & 0 & 0 & 0\\
v & E & v & 0 & 0 & 0 & 0 & 0 & 0 & 0 & 0\\
0 & v & 2E & v & 0 & 0 & 0 & 0 & 0 & 0 & 0\\
0 & 0 & v & 3E & v & 0 & 0 & 0 & 0 & 0 & 0\\
0 & 0 & 0 & v & 4E & v & 0 & 0 & 0 & 0 & 0\\
0 & 0 & 0 & 0 & v & 5E & v & 0 & 0 & 0 & 0\\
0 & 0 & 0 & 0 & 0 & v & 6E & v & 0 & 0 & 0\\
0 & 0 & 0 & 0 & 0 & 0 & v & 7E & v & 0 & 0\\
0 & 0 & 0 & 0 & 0 & 0 & 0 & v & 8E & v & 0\\
0 & 0 & 0 & 0 & 0 & 0 & 0 & 0 & v & 9E & v\\
0 & 0 & 0 & 0 & 0 & 0 & 0 & 0 & 0 & v & 10E
\end{bmatrix}
\end{gather*}

Note that along the diagonal we have equally spaced energies E$_{n}$=
n*E. There is only one coupling parameter, v, and the coupling is only
to the nearest neighbors.

We choose E to be 1 MeV. We will consider four choices of the coupling,
v=0.5 (weak), v=1, v=2, and finally v=3 (strong). (The relevant parameter
is really v/E)

The ``wave functions'' are column vectors with 11 entries ($a_{1}$, a$_{2}$,....., a$_{11}$).
We have a ground state and 10 excited states.

\section{The distributions}

The strength matrix element O between a state \{a\} and a state \{b\}
is simply

\begin{equation}
O=(a_{1}b_{2}+.....+a_{10}b_{11}) + (b_{1}a_{2}+.....+b_{11}a_{10})\textless{}\textless{}n T (n+1)\textgreater{} \textgreater{}.
\end{equation}

As mentioned above,we take \textless{}\textless{}nT(n+1)\textgreater{}\textgreater{}
to be 1. The Strength is O$^{2}$ and we plot ln(O$^{2}$).

\begin{figure}
\centering \includegraphics[width=0.7\linewidth]{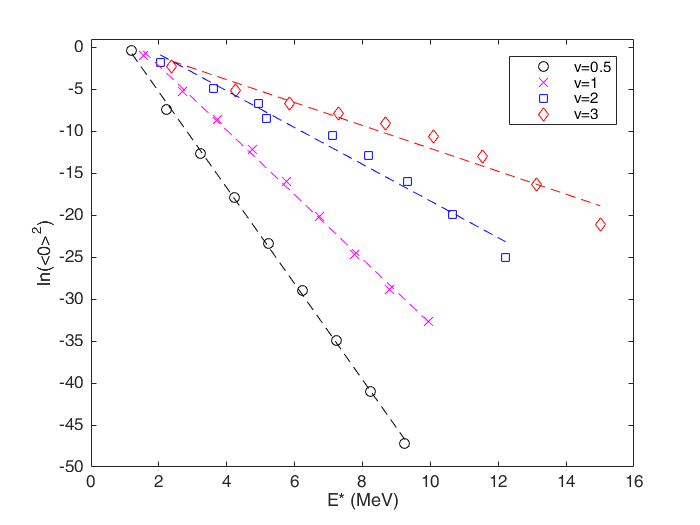}
\caption{Log of the Strength vs. Excitation Energy}
\label{fig1} 
\end{figure}

\FloatBarrier

\section{Special Features of our model.}

There are some special features of our model. First of all, if we
change the coupling v to -v we get the same results. This can be understood
easily. Suppose we multiply every other basis state by a minus sign
(second, fourth, sixth, etc.). The negative of a wave function is
the same as the original wave function. What this does in effect is
change the coupling v to -v. But, since we are not really changing the
basis wave functions we expect the same results as before.

One consequence of this is that the absolute value of the energy of
the fifth excited state is 5 MeV no matter what the value of the coupling
strength is. This is connected with the fact that the fifth is the
middle state with 5 states pushing up on it from below and five from
above.

Another perhaps more interesting fact is that there is a simple relation
between the eigenfunctions of the lowest state and the highest state.
If the lowest eigenstate is of the form
$(a_{1}$, $a_{2}$, $a_{3}$, $a_{4}$.....$a_{11}$) then the
highest one has as an eigenfunction $(a_{11}, -a_{10}, a_{9},
-a_{8}.....a_{2},-a_{1})$. There are similar relationships
between the second lowest and second highest states etc.. This has 
the consequence that the strength matrix element between
the lowest state and the highest state vanishes no matter what the
coupling strength is. This is convenient because when things become
very small they test the precision of the calculations. Knowing this
we do not include the low-high transitions in our fits.

\section{Discussion of results }

In Figure 1 we show a log plot (base e) of the strength from the ground
state to the 10 excited states. If the behavior is that of a decreasing
exponential then in a log plot we should see a straight line with
a negative slope. We performed linear fits to ln(\textless{}O\textgreater{}$^{2}$)
vs excitation energy, i.e. ln (\textless{}O\textgreater{}$^{2}$)
=a +b E{*}. As mentioned above we do not include the transition from
the lowest state to the highest state because in that case \textless{}O\textgreater{}
must have a zero value.

The linear fits for various v's are as follows:

v=0.5: 7.38-5.72E{*}

v=1: 5.44-3.84E{*}

v=2: 3.60-2.20E{*}

v=3: 1.42-1.34E{*}

Indeed, in the log plots in Figure 1 corresponding to v=0.5, v=1, v=2
and v=3 the dominant behavior is of a linear decrease, hence an exponential
decrease in strength.

We should however not expect an exponential behavior in all cases.
Considering the extreme situation where there is no coupling (v=0).In
that case there would only be one transition--from the ground to the
first excited state. This would look like a spike in a plot of strength
vs excitation energy - very far from a decreasing exponential. However
even with the smallest v=0.5 shown here we get to a good approximation
of an exponential decrease.

The standard deviation for v=0.5 is 0.2001. For v=1 it is 0.2089. For
v=2 it is 0.5379 and for v=3 it is 0.5683. To a first approximation
the results for all these cases show an exponential fall-off. 

Note that the magnitude of the slope decreases with increasing v.
This could be understood from the fact that when v=0 one only reaches
the first excited state. There is no strength to higher states. As
we increase v we increase the mixing of the basis states and so it
is more favorable to reach states at higher energy. 

The values of\textless{}\textbar{}O$^{2}$\textbar{}\textgreater{}
for transitions from ground to the first excited state , for v=0, 0.5, 1, 2, and
3 are respectively 1, 0.6838, 0.3833, 0.2195, and 0.0983.

\section{Degenerate energies. }

What happens if in the Hamiltonian we take E to be zero, i.e. all the
energies along the diagonal are the same? Then we only have off diagonal
v's. A very interesting phenomenon occurs. All transitions rates,
as defined by Eq 1 in Sec 4 from any state to any other state vanish.
This can be easily explained. When we have all zeros on the diagonal
of our Hamiltonian, the Hamiltonian matrix becomes proportional to
the transition operator, represented as a matrix. The Hamiltonian,
acting on an eigenstate gives back the same eigenstate and hence there
will be no transition.

One thus verifies that all transition matrix elements are zero. One
can however have diagonal non-vanishing matrix elements.

\section{Other matrix models}

Our contention that matrix models can be insightful is strengthened
by the fact that there are examples in the literature where this is
the case. We cite examples where the couplings in the matrices are
different from ours and the problems that are addressed are also different.

Consider an n by n matrix M in which all elements are the same, say
c. Then it is easy to show that

{\centering
$M^{2} = cnM$ or $M(1-cnM) = 0$\par
}

There are n-1 degenerate states with energy zero and one ``collective
state'' with energy nc. For c positive the collective state is at
a high energy and can have an association with isovector modes like
the giant dipole state. For c negative the collective state is at
a lower energy than the degenerate states. One common association
in this case is with isoscalar octupole states. Note the difference
with our model where, as we mention above, we get the same result
when we change the sign of the coupling.

Bohr and Mottelson{[}10{]} used a matrix model to present an alternate
derivation of the Breit-Wigner formula{[}11{]} for the strength function
of a resonance. In their words ``We wish to describe how the amplitude
for a particular channel a may be distributed over many stationary
states of a complicated system.'' Their coupling is from a single
particle state a to nearby complicated states and in one example they
have a constant coupling to all these states. Their coupling causes
the single particle sharp state to get broadened out to a resonance
state with a width $\varGamma$. In the model considered here we only
have couplings to nearest neighbor states, and the exponential behavior
we show in Fig 1. is drastically different from the Breit-Wigner
resonance shape. Still it is nice to see that matrix models can be
useful in a wide variety of problems.

\section{Closing remarks }
We have made a plausible case here that an exponential decrease
of strength, although not universal, is widespread. Except for the
case where the coupling is very weak we will get such a decrease. We
find this to be the case with the simple matrix we have constructed
and also in realistic calculations of magnetic dipole transitions
discussed in refs.{[}4,5{]}. In the future we will consider other
matrix models and strength distributions for other operators.

\end{document}